\documentclass[aps,reprint,prl,10pt,twocolumn,showkeys,nofootinbib,showpacs,superscriptaddress]{revtex4-1}
\usepackage{graphicx}
\usepackage{url}
\usepackage{amsmath}
\usepackage{amssymb}
\usepackage{bm}
\usepackage{dsfont}
\usepackage{subfigure}
\usepackage{dcolumn}
\usepackage{braket}
\usepackage{bm}
\usepackage{bbm}
\usepackage{color}
\usepackage{pdfpages}
\usepackage{pgffor}
\usepackage{animate}
\usepackage{physics}

\usepackage{notoccite}

\makeatletter
\AtBeginDocument{\let\LS@rot\@undefined}
\makeatother

\newcommand{\Li}{$^{6}\mathrm{Li}\;$}

\newcommand{\affcua}{MIT-Harvard Center for Ultracold Atoms, Research Laboratory of Electronics, and Department of Physics,
Massachusetts Institute of Technology, Cambridge, Massachusetts 02139, USA}

\newcommand{\affens}{D\'{e}partement de Physique, Ecole Normale Sup\'{e}rieure / PSL Research University, CNRS, 24 rue Lhomond, 75005 Paris, France}

\begin{document}

\title{Spectral Response and Contact of the Unitary Fermi Gas}
\author{Biswaroop Mukherjee}
\affiliation{\affcua}
\author{Parth B. Patel}
\affiliation{\affcua}
\author{Zhenjie Yan}
\affiliation{\affcua}

\author{\\ Richard J. Fletcher}
\affiliation{\affcua}
\author{Julian Struck}
\affiliation{\affcua}
\affiliation{\affens}
\author{Martin W. Zwierlein}
\affiliation{\affcua}

\begin{abstract}
We measure radio frequency (rf) spectra of the homogeneous unitary Fermi gas at temperatures ranging from the Boltzmann regime through quantum degeneracy and across the superfluid transition. For all temperatures, a single spectral peak is observed. Its position smoothly evolves from the bare atomic resonance in the Boltzmann regime to a frequency corresponding to nearly one Fermi energy at the lowest temperatures. At high temperatures, the peak width reflects the scattering rate of the atoms, while at low temperatures, the width is set by the size of fermion pairs. Above the superfluid transition, and approaching the quantum critical regime, the width increases linearly with temperature, indicating non-Fermi-liquid behavior. From the wings of the rf spectra, we obtain the contact, quantifying the strength of short-range pair correlations. We find that the contact rapidly increases as the gas is cooled below the superfluid transition.

\end{abstract}


\maketitle

Understanding fermion pairing and pair correlations is of central relevance to strongly interacting Fermi systems such as nuclei~\cite{Hen2014, Weiss2015}, ultracold gases~\cite{Inguscio2008, Giorgini2008, Zwerger2016, Zwierlein2015}, liquid $^3$He~\cite{Leggett1972}, high temperature superconductors~\cite{Lee2006}, and neutron stars~\cite{Baym1969}. Strong interactions on the order of the Fermi energy challenge theoretical approaches, especially methods that predict dynamic properties such as transport or the spectral response at finite temperature~\cite{Enss2012}. Atomic Fermi gases at Feshbach resonances realize a paradigmatic system where the gas becomes as strongly interacting as allowed by unitarity~\cite{Inguscio2008, Giorgini2008, Zwerger2016, Randeria2014, Zwierlein2015}. Here, the system becomes universal, requiring only two energy scales: the Fermi energy $E_F$ and thermal energy $k_B T$, where $k_B$ is the Boltzmann constant and $T$ is the temperature. The corresponding length scales are the interparticle spacing $\lambda_F\,{=}\, n^{-1/3}$ and the thermal de Broglie wavelength $ \lambda_{ T}\,{=}\,h/\sqrt{2\pi mk_BT}$, where $m$ and $n$ are the mass and number density of the atoms respectively. When the two energy scales are comparable, the system enters a quantum critical regime separating the high temperature Boltzmann gas from the fermionic superfluid~\cite{Nikolic2007}. Quantum criticality is often associated with the absence of quasiparticles~\cite{Nikolic2007,  Enss2012, Frank2018}, spurring a debate on the applicability of Fermi liquid theory to the degenerate normal fluid below the Fermi temperature $T_F = E_F/k_B$ but above the superfluid transition temperature $T_c\approx 0.167\,T_F$~\cite{Ku2012, Nascimbene2011b, Rothstein2019}. It has been conjectured that preformed pairs exist above $T_c$, up to a pairing temperature $T^*$~\cite{Inguscio2008, Zwerger2016, Gaebler2010, Sagi2015, Perali2002a, Hu2010, Magierski2011, Randeria2014}.

Radio frequency (rf) spectroscopy measures the momentum integrated, occupied spectral function, providing a powerful tool for studying interactions and correlations in Fermi gases~\cite{Gupta2003, Regal2003a, Regal2003, Chin2004, Schunck2008b, Haussmann2009}. Here, a particle is ejected from the interacting many-body state and transferred into a weakly interacting final state. Shifts in rf spectra indicate attractive or repulsive interactions in the gas. At high temperatures, the width of the rf spectrum reflects the scattering rate in the gas, while at low temperatures, the width has been used to infer the pair size of superfluid fermion pairs~\cite{Schunck2008b}. 

The high frequency tails of the rf spectra are sensitive to the spectral function at high momenta, and therefore are governed by short range correlations quantified by the contact, which also determines the change of the energy with respect to the interaction strength~\cite{Tan2008, Tan2008c, Tan2008d}. From the momentum distribution within nuclei~\cite{Hen2014, Weiss2015} to the frequency dependence of the shear viscosity in ultracold fermionic superfluids~\cite{Taylor2010, Enss2011b}, the contact is central to Fermi gases dominated by short-range interactions. Since the contact is proposed to be sensitive to the superfluid pairing gap, it could signal a pseudogap regime above $T_c$~\cite{Pieri2009, Palestini2010a, Enss2011b, Mueller2017}. Although the temperature dependence of the contact near $T_c$ has been the subject of many theoretical predictions, a consensus has not been reached~\cite{Hu2011,Enss2011b,Goulko2016,Rossi2018}.

\begin{figure*}[!t]
\centering
\includegraphics[width=\textwidth]{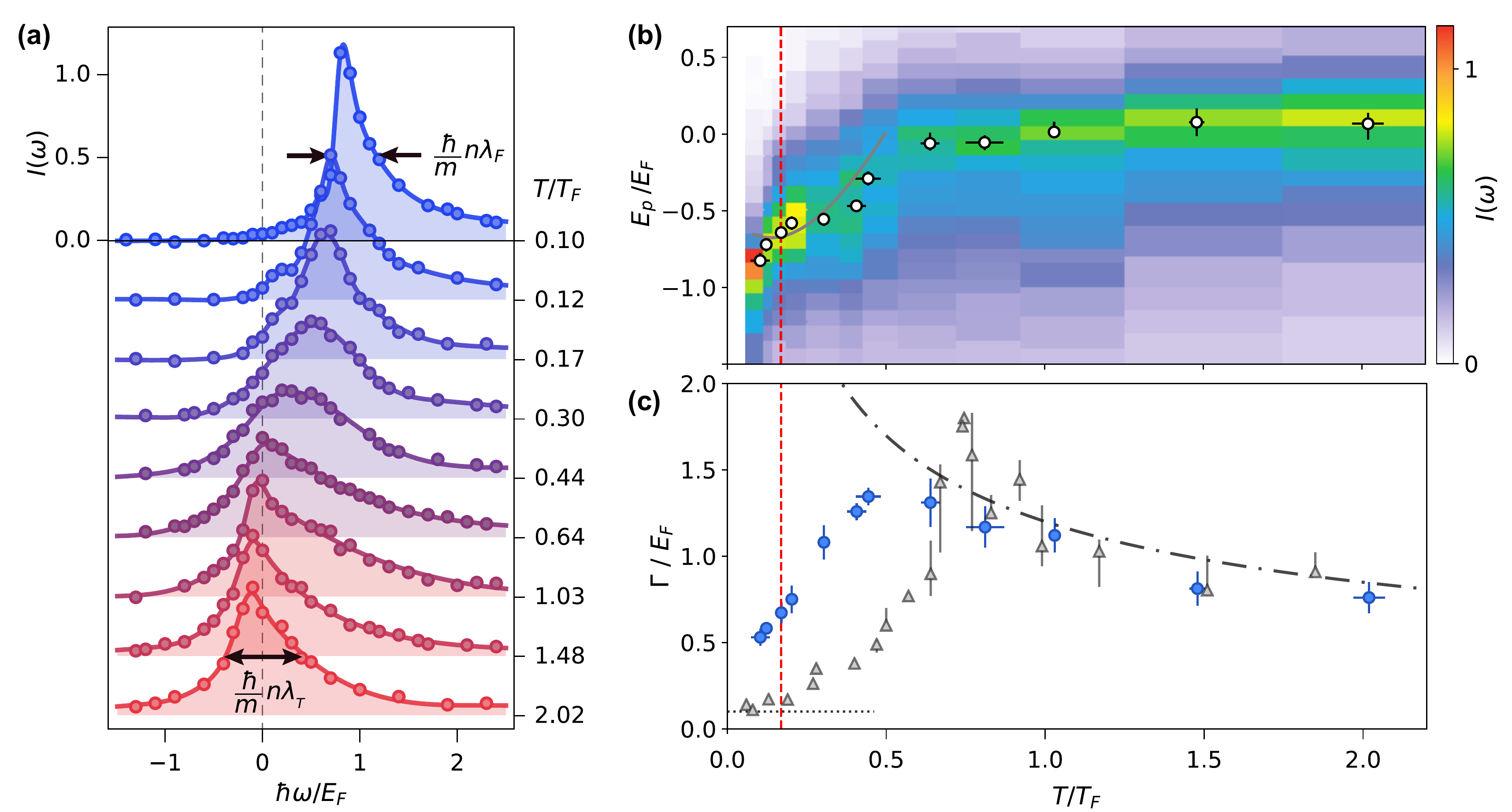}
\caption{  (a)  Thermal evolution of rf spectra. The Rabi frequency is $\Omega_R = 2\pi\times0.5$ kHz and the pulse duration is $T_{\mathrm{Pulse}}\,{=}\,1$~ms.  The solid lines are guides to the eye. (b) Frequency of the peak ($E_p=-\hbar\omega$) of the rf spectra as a function of temperature shown as white dots on an intensity plot of the rf response. The grey solid line is a solution to the Cooper problem at nonzero temperature~\cite{suppmat}. (c) The full width at half maximum $\Gamma$ of the rf peak as a function of $T/T_F$. The black dotted-dashed line $\Gamma/E_F = 1.2 \sqrt{T_F/T}$ shows the temperature dependence of the width due to scattering in the high-temperature gas~\cite{Enss2011b,Sun2015}. The grey triangles are the corresponding width measurements of a highly spin-imbalanced gas~\cite{Yan2018}. The horizontal black dotted line represents the Fourier broadening of $0.1\,E_F$~\cite{suppmat}. The vertical dashed red line in both (b) and (c) marks the superfluid transition~\cite{Ku2012}. 
}
\label{fig:M1}
\end{figure*}

 Initial studies of unitary Fermi gases using rf spectroscopy were affected by inhomogeneous densities in harmonic traps, yielding doubly peaked spectra that were interpreted as observations of the pairing gap~\cite{Chin2004, Schunck2007}, and from the influence of interactions in the final state, which caused significantly narrower spectra and smaller shifts than expected~\cite{Gupta2003, Schunck2007, Baym2007a, Punk2007}. Measurements of the contact, made using both rf~\cite{Stewart2010a, Shkedrov2018} and Bragg~\cite{Kuhnle2010, Kuhnle2011, Hoinka2013} spectroscopy, were also broadened by inhomogeneous potentials. To avoid trap broadening, tomographic techniques have been used to measure local rf spectra, yielding measurements of the superfluid gap~\cite{Schirotzek2008}, the spectral function~\cite{Gaebler2010, Sagi2015}, and the contact~\cite{Sagi2012}.  A recent advance has been the creation of uniform box potentials~\cite{Gaunt2013, Mukherjee2017b, Hueck2018}. These are ideal for rf spectroscopy and precision measurements of the contact: since the entire cloud is at a constant density, global probes such as rf address all atoms, and benefit from a stronger signal.

In this Letter, we report on rf spectroscopy of the homogeneous unitary Fermi gas in a box potential. A single peak is observed for all temperatures from the superfluid regime into the high temperature Boltzmann gas. The tails of the rf spectra reveal the contact, which shows a rapid rise as the temperature is reduced below $T_c$.


We prepare~\Li atoms in two of the three lowest hyperfine states $\ket{\downarrow}{=}\ket{1}$ and $\ket{\uparrow}{=}\ket{3}$ at a magnetic field of 690~G, where interspin interactions are resonant. A uniform optical box potential with cylindrical symmetry is loaded with $N\,{\sim}\,10^6$ atoms per spin state (with Fermi energies $E_F\,{\sim}\,h\,{\times}\,10$~kHz), creating spin-balanced homogeneous gases at temperatures ranging from $T/T_F\,{=}\,0.10$ to $3.0$~\cite{Mukherjee2017b}. A square rf pulse transfers atoms from state $\ket{\downarrow}$ into state $\ket{f} = \ket{2}$. Final state interactions between atoms in state $\ket{f}$ and atoms in states $\ket{\uparrow}$ and $\ket{\downarrow}$ are small ($k_Fa_{f}\,{\lesssim}\,0.2$, where $a_f$ is the scattering length characterizing collisions between atoms in the final and initial states, and $\hbar k_F\,{=}\,\sqrt{2mE_F}$ is the Fermi momentum)~\cite{Schunck2008b}. After the rf pulse, we measure the atom numbers $N_\downarrow$ and $N_f$ in the initial and final states. Within linear response, according to Fermi's golden rule, $N_f$ is proportional to the pulse time $T_{\mathrm{Pulse}}$, the square of the single-particle Rabi frequency $\Omega_R$ and an energy density of states. Thus, we define a normalized, dimensionless rf spectrum as $I(\omega)\,{=}\,(N_f(\omega)/N_\downarrow)(E_{F}/\hbar\Omega_R^2T_{\mathrm{Pulse}})$~\cite{suppmat,Yan2018}. Because of the scale invariance of the balanced unitary Fermi gas, this dimensionless function can only depend on $T/T_F$ and $\hbar \omega/E_F$.

For thermometry, we release the cloud from the uniform potential into a harmonic trap along one direction~\cite{Yan2018}. Since the cloud expands isoenergetically, the resulting spatial profile after thermalization provides the energy per particle, which can be related to the reduced temperature, $T/T_F$, using a virial relation and the measured equation of state~\cite{Ku2012}. To clearly identify the superfluid transition, we measure the pair momentum distribution by a rapid ramp of the magnetic field to the molecular side of the Feshbach resonance before releasing the gas into a harmonic trap for a quarter period~\cite{Mukherjee2017b, suppmat}.

Initially, we focus on changes in the line shape for rf frequencies within ${\sim}E_F/\hbar$ of the bare (single-particle) resonance (see Fig~\ref{fig:M1}(a)), and follow the changes in the peak position $E_p$ (shown in Fig~\ref{fig:M1}(b)).  As the hot spin-balanced Fermi gas is cooled below the Fermi temperature, the peak shift decreases from roughly zero for temperatures $T\,{\gtrsim}\,T_F$, to $E_p\,{\approx}\,-0.8~E_F$ for temperatures below the superfluid transition temperature (see Fig.~\ref{fig:M1}(b)). At high temperatures, one might na\"{\i}vely expect a shift on the order of $E_p\sim \hbar n \lambda_T/m$ due to unitarity-limited interactions in the gas. However, there exists both an attractive and a repulsive energy branch, which are symmetric about zero at unitarity~\cite{Ho2004}, and when $T\,\gg\,T_F$, their contributions to the shift cancel~\cite{Enss2011b,Sun2015, Fletcher2017}. As to the interpretation of the peak shift at degenerate temperatures, a solution to the Cooper problem in the presence of a $T>0$ Fermi sea shows that it is energetically favorable to form pairs when $T\lesssim0.5\,T_F$~\cite{suppmat}, and the resulting pair energy agrees qualitatively with the observed shifts (grey line in Fig.~\ref{fig:M1}(b)). However, it is known that fluctuations suppress the onset of pair condensation and superfluidity to $0.167(13)\,T_F$~\cite{Nozieres1985, Ku2012, Randeria2014, Zwerger2016}. In a zero-temperature superfluid, BCS theory would predict a peak shift given by the pair binding energy $E_B = \Delta^2/2 E_{  F}$, where $\Delta$ is the pairing gap~\cite{Inguscio2008}. Including Hartree terms is found to result in an additional shift of the peak~\cite{Schirotzek2008,Haussmann2009}.


Now, we turn to the widths, $\Gamma$, defined as the full width at half maximum of the rf spectra (see Fig.~\ref{fig:M1}(c)). As the gas is cooled from the Boltzmann regime, the width gradually increases, and attains a maximum of $\Gamma\,{=}\,1.35(5)\,E_\mathrm{F}$ near $T\,{=}\,0.44(4)\,T_F$. For temperatures much higher than $T_F$, the system is a Boltzmann gas of atoms scattering with a unitarity limited cross section $\sigma \sim \lambda_{  T}^2$. Transport properties and short-range pair correlations are governed by the scattering rate $\Gamma = n_ \downarrow \sigma \langle v_{\rm rel}\rangle \sim \hbar n_ \downarrow  \lambda_{  T}/m$ and a mean-free path $l = (n_ \downarrow \sigma)^{-1} \sim (n_ \downarrow \lambda_{  T}^2)^{-1}$, where $n_ \downarrow$ is the density of atoms in $\left| \downarrow\right>$, and $\langle v_{\rm rel}\rangle \sim \hbar /m \lambda_{  T}$ is the thermally averaged relative velocity. This leads to a width that scales as $\Gamma\propto1/\sqrt{T}$, shown as the dotted-dashed line in Fig.~\ref{fig:M1}(c)~\cite{Enss2011b}.

As the cloud is cooled below $T\,{\approx}\,0.5\,T_F$, the width decreases linearly with temperature to $\Gamma\sim0.52\,E_\mathrm{F}/\hbar$ in the coldest gases measured ($T=0.10(1)\,T_\mathrm{F}$). For temperatures below $T_c$, we expect the gas to consist of pairs of size $\xi$. The rf spectrum will be broadened by the distribution of momenta $\sim \hbar/\xi$ inside each pair, leading to a spread of possible final kinetic energies $\hbar^2 k^2/m \sim \hbar^2/m \xi^2$ and a corresponding spectral width $\hbar/m\xi^2$. At unitarity and at $T=0$, the pair size is set by the interparticle spacing $\lambda_F$~\cite{Zwerger2016, Inguscio2008, Schunck2008b}. Thus the rf width at low temperatures is $\Gamma \sim  \hbar n \lambda_{  F}/m$.

For temperatures above $T_c$, it has been suggested that the normal fluid can be described as a Fermi liquid~\cite{Nascimbene2010, Nascimbene2011b}. This would imply a quadratic relation between the peak width and the temperature~\cite{Nozieres1966}, as observed in the widths of the rf spectra of Fermi polarons at unitarity~\cite{Yan2018}. However, the measured width of the spin-balanced Fermi gas changes linearly in temperature, implying non-Fermi liquid behavior in the normal fluid. In addition, $\Gamma>E_F/\hbar$ for $0.3\,{\lesssim}\, T/T_F\, {\lesssim}\,1.2$, indicating a breakdown of well-defined quasiparticles over a large range of temperatures near the quantum critical regime~\cite{Nikolic2007, Enss2012, Frank2018}.


\begin{figure}[]
\centering
\includegraphics[width=8.6cm]{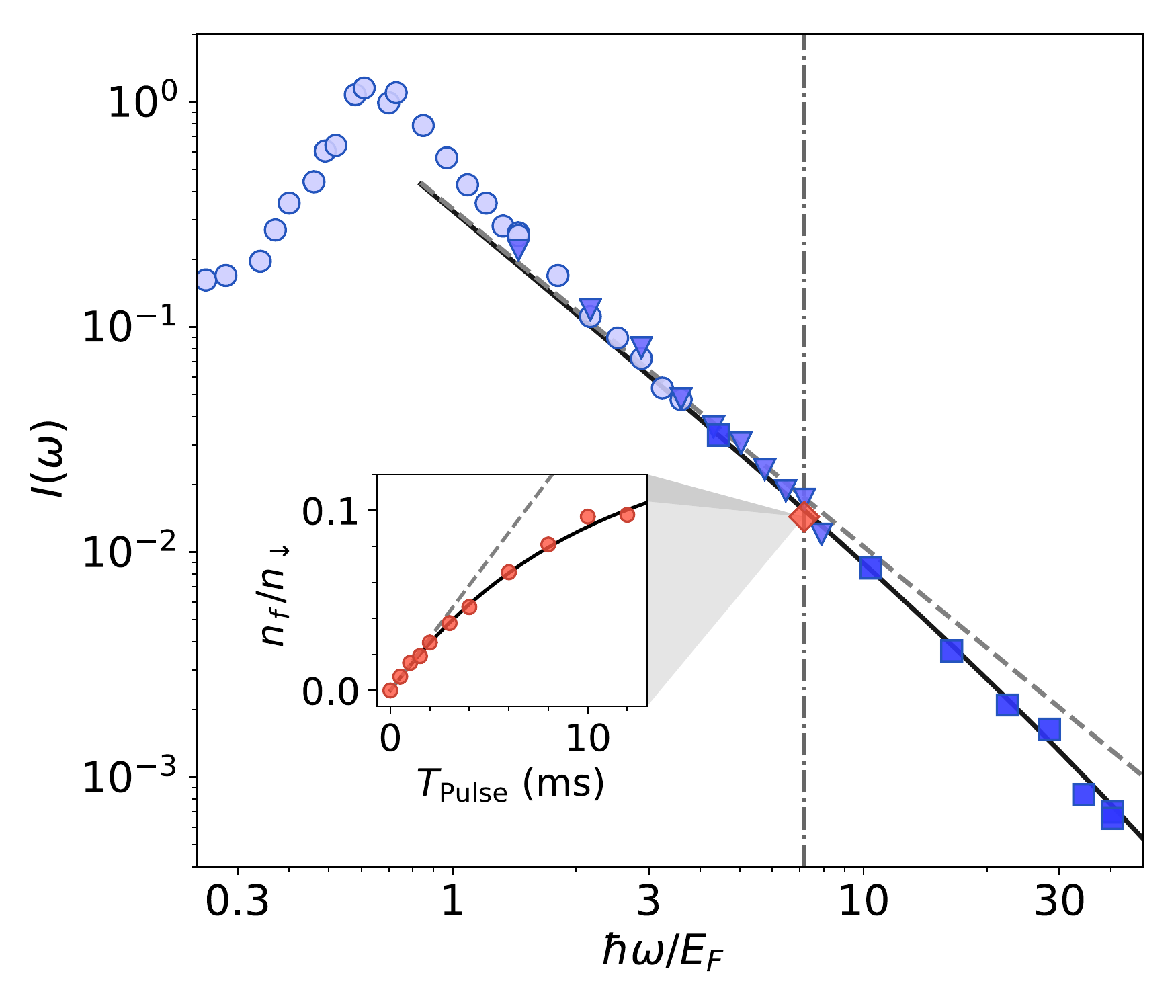}
\caption{Rf spectrum at high frequencies. Here, the temperature of the gas is $T/T_F\,{=}\,0.10(1)$, the pulse duration is $T_\mathrm{Pulse}\,{=}\,1$ ms, and the Rabi frequencies are $2\pi\,{\times}\,536$ Hz (light blue circles), $2\pi\,{\times}\,1.20$~kHz (medium blue triangles), and $2\pi\,{\times}\,3.04$ kHz (dark blue squares). The black solid line shows a fit of Eq.~\ref{finalstaterf} to the data, while the grey dashed line shows the fit neglecting final state interactions. The contact can be directly obtained from the transfer rate at a fixed detuning of $60 $ kHz ($\hbar\omega/E_F\sim7.1$) (dotted-dashed vertical line). Inset: we vary the pulse time at this fixed detuning, and extract the initial slope (dashed line) of the exponential saturating fit (solid line). The rf transfer rate obtained from the initial linear slope is shown as the red diamond in the main plot. Here, $\Omega_R =2\pi\times1.18$ kHz.}
\label{fig:M2}
\end{figure}

We now consider the rf spectrum at frequencies much larger than $E_F/\hbar$, where the rf-coupled high-momentum tails reveal information about the short-range pair correlations between atoms. In a gas with contact interactions, the pair correlation function at short distances is $ \lim_{r\rightarrow0} \langle n_\uparrow({\bf{r_0}+{\bf r/2}})n_\downarrow({\bf{r_0}}-{\bf{r/2}})\rangle= C/(4\pi r)^2$. The contact $C$ connects a number of fundamental relations, independent of the details of the short-range interaction potential~\cite{Tan2008}. In particular, the contact governs the momentum distribution at large momenta: $\lim_{k\rightarrow\infty}n(k) = C/k^4$. For rf spectroscopy, the density of final states scales as $\sqrt{\omega}$, and the energy cost to flip a spin at high momenta is $\lim_{k\rightarrow\infty}\hbar\omega=\hbar^2k^2/m$. Thus, the number of atoms transferred by the rf pulse at high frequencies in linear response is $\propto C/\omega^{3/2}$~\cite{Haussmann2009, Zwerger2016}. Including final state interactions, the general expression for the rf transfer rate in a gas with unitarity-limited initial state interactions is~\cite{Braaten2010}:
\begin{equation}
\lim_{\omega \rightarrow \infty}I({\omega})\,{=}\left( \frac{C}{N k_{\mathrm{F}}} \right)\frac{1}{2\sqrt{2}\pi\,(1+ \hbar \omega/E_{b})} \left(\frac{E_{\mathrm{F}}}{\hbar \omega} \right)^{3/2},
\label{finalstaterf}
\end{equation}
where $N = N_\uparrow + N_\downarrow$ is the total number of atoms, and the final state molecular binding energy is $E_b\,{=}\, \hbar^2/ma_{f}^2\,{\approx}\,h\,{\times}\,433$ kHz $\approx40\,E_F$.   Figure~\ref{fig:M2} shows a typical rf spectrum at $T/T_F=0.10$, with a fit of Eq.~\ref{finalstaterf} to data with detunings $\hbar\omega>3\,E_F$, using the dimensionless contact $\tilde{C} = C/Nk_{\mathrm{F}}$ as the only free parameter. At detunings larger than about 10~$E_F$, the data deviate from a typical $\omega^{-3/2}$ tail, and is better described by the full expression Eq.~\ref{finalstaterf} including final state interactions. Here, the Rabi frequency was varied across the plot to ensure small transfers near the peak and a high signal-to-noise ratio at detunings up to $\hbar\omega/E_F = 31$. The fit of Eq.~\ref{finalstaterf} to the data gives a low-temperature contact of $\tilde{C} = 3.07(6)$, consistent with a quantum Monte Carlo result $\tilde{C} = 2.95(10)$~\cite{Drut2011}, the Luttinger-Ward (LW) calculation $\tilde{C}=3.02$~\cite{Haussmann2009}, as well as previous measurements using losses $\tilde{C}=3.1(3)$~\cite{Laurent2017} and Bragg spectroscopy $\tilde{C}=3.06(8)$~\cite{Hoinka2013}.

For a more efficient measurement of the contact across a range of temperatures, we vary the pulse time at a fixed detuning of $60$ kHz ($\hbar\omega/E_F \gtrsim 6$) that is large compared to the Fermi energy and temperature.~\cite{suppmat}. Deviations from linear response are observed for transfers as small as 5\% (see inset of Fig.~\ref{fig:M2}). We fit the transfers to an exponentially saturating function $ A(1-\exp(-T_\mathrm{Pulse}/\tau))$, and find the initial linear slope $A/\tau$ in order to extract the contact for each temperature using Eq.~\ref{finalstaterf}. This ensures that every measurement is taken in the linear response regime. 

\begin{figure}[!t]
\centering
\includegraphics[width=8.6cm]{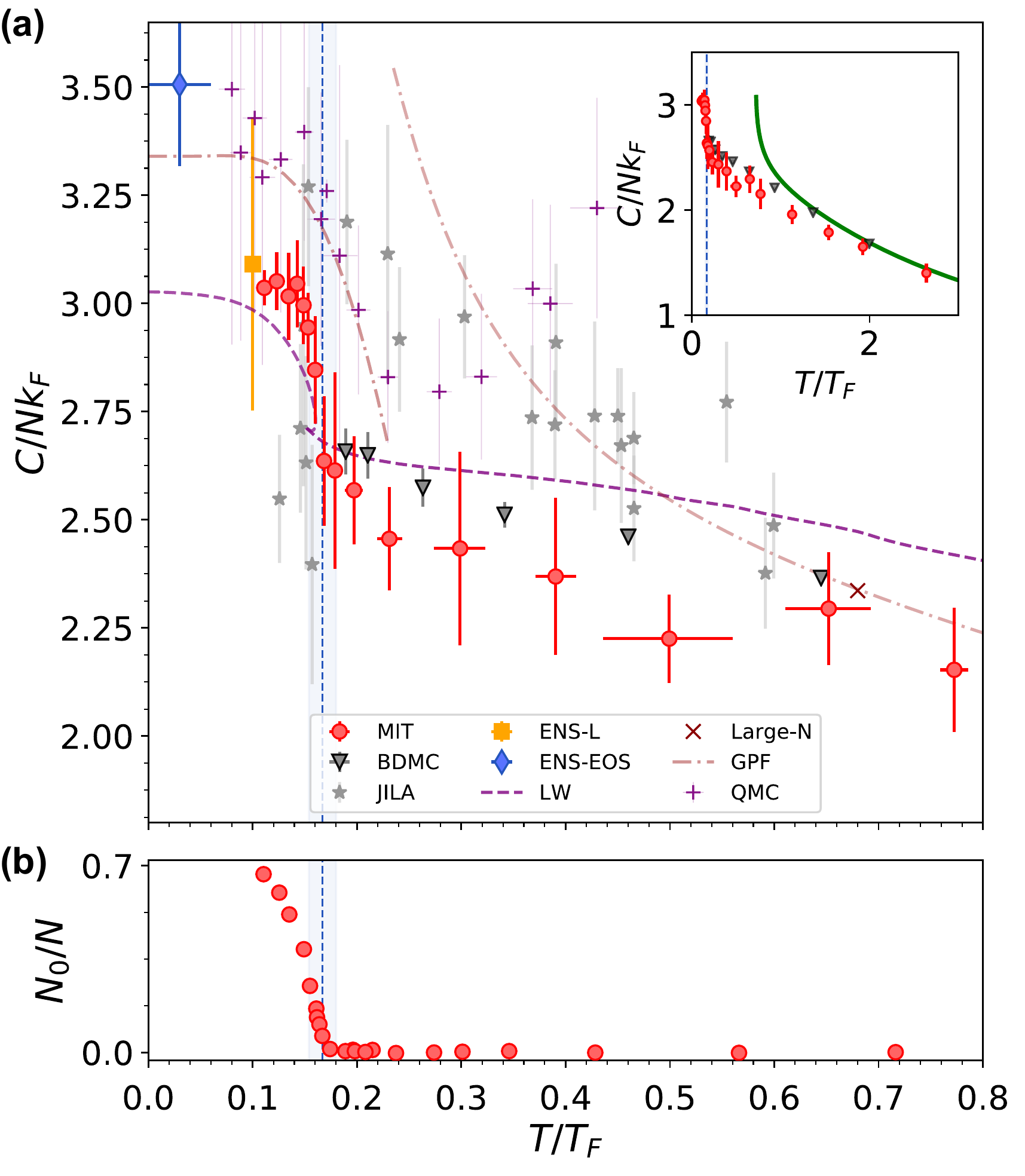}
\caption{The dimensionless contact $C/Nk_F$~(a) and condensate fraction $N_0/N$~(b) of the unitary Fermi gas as a function of the reduced temperature $T/T_F$. Our measurements of the contact (red points) are compared with a number of theoretical estimates: bold-diagrammatic Monte Carlo (BDMC)~\cite{Rossi2018}, QMC~\cite{Goulko2016}, Luttinger-Ward (LW)~\cite{Enss2011b}, large-N~\cite{Enss2012}, and Gaussian pair fluctuations (GPF)~\cite{Hu2011}. Also shown is the homogeneous contact obtained from the equation of state (ENS-EOS)~\cite{Nascimbene2010}, from loss rate measurements (ENS-L)~\cite{Laurent2017}, and from rf spectroscopy by the JILA group~\cite{Sagi2015} across a range of temperatures. The vertical blue dotted lines and light blue shaded vertical regions mark $T_c/T_F\,{=}\,0.167(13)$~\cite{Ku2012}. The inset of (a) shows the contact over a wider range of temperatures and marks the high-temperature agreement with the third order virial expansion. The error bars account for the statistical uncertainties in the data.}
\label{fig:M3}
\end{figure}

In Fig.~\ref{fig:M3}(a), we show the temperature dependence of the contact. As the gas is cooled, the contact shows a gradual increase down to the superfluid transition $T_c$. Entering the superfluid transition, the contact rapidly rises by approximately 15\%. The changes in the contact reveal the temperature dependence of short-range pair correlations in the spin-balanced Fermi gas. At temperatures far above $T_F$, the contact reflects the inverse mean free path in the gas $1/l\sim 1/T$. At lower temperatures, the behavior of the contact is better described by a third-order virial expansion (see inset of ~\ref{fig:M3}(a))~\cite{Hu2011}. Near $T_c$, predictions of the contact vary considerably. In the quantum critical regime, a leading-order 1/$N$ calculation (equivalent to a Gaussian pair fluctuation or Nozi\`eres--Schmitt-Rink method) results in a prediction $\tilde{C}(\mu=0, T\approx0.68~T_F) = 2.34$~\cite{Enss2012}, which is consistent with our measurement of $\tilde{C}(T= 0.65(4)~T_F) = 2.29(13)$. For temperatures above the superfluid transition, our data agree well with both a bold diagrammatic Monte Carlo calculation~\cite{Rossi2018}, and, especially near $T_c$, the LW calculation~\cite{Enss2011b}. The contact rises as the temperature is decreased below $T_c$, a feature captured by the LW formalism, in which the contact is directly sensitive to pairing: $\tilde{C}\sim (\Delta/E_F)^2$~\cite{Pieri2009, Haussmann2009}. While short-range pair correlations do not necessarily signify pairing~\cite{Mueller2017}, the rapid rise of the contact below $T_c$ is strongly indicative of an additional contribution from fermion pairs, as predicted by LW. At temperatures $T\ll T_c$, below the reach of our experiment, phonons are likely the only remaining excitations in the unitary Fermi gas, and are expected to contribute to the contact by an amount that scales as $T^4$~\cite{Yu2009a}.


In conclusion, rf spectroscopy of the homogeneous unitary Fermi gas reveals strong attractive interactions, the non-Fermi-liquid nature of excitations in the gas across the quantum critical regime, and a rapid increase in short-range pair correlations upon entering the superfluid regime. The strong variation with temperature of the position of the spectral peak may serve as a local thermometer in future studies of heat transport in ultracold Fermi gases. Furthermore, these measurements of the contact provide a benchmark for many-body theories of the unitary Fermi gas. The uniform trap enables direct access to homogeneous measurements of thermodynamic quantities, and increases sensitivity to abrupt changes of those quantities near phase transitions. This could be particularly useful in the limit of high spin imbalance, where the nature of impurities suddenly transitions from Fermi polarons to molecules~\cite{Punk2009, Schirotzek2009}.  

We note that measurements of the temperature dependence of the contact were simultaneously performed at Swinburne using Bragg spectroscopy~\cite{Carcy2019}. Their data are in excellent agreement with the present results.

We thank C. J. Vale, F. Werner, and W. Zwerger for helpful discussions. This work was supported by the NSF, AFOSR, ONR, the AFOSR MURI on Exotic Phases, and the David and Lucile Packard Foundation. J.S. was supported by LabEX ENS-ICFP: ANR-10-LABX-0010/ANR-10-IDEX-0001-02 PSL*.



%

\clearpage


\section*{Supplementary material}

# Supplemental Material:
# Spectral response and contact of the unitary Fermi gas

Biswaroop Mukherjee,[1] Parth B. Patel,[1] Zhenjie Yan,[1]
Richard J. Fletcher,[1] Julian Struck,[1,2] and Martin W. Zwierlein[1]

[1]*MIT-Harvard Center for Ultracold Atoms, Research Laboratory of Electronics, and Department
of Physics, Massachusetts Institute of Technology, Cambridge, Massachusetts 02139, USA*
[2]*Département de Physique, Ecole Normale Supérieure / PSL Research University, CNRS, 24 rue Lhomond, 75005 Paris, France*

## DENSITY CALIBRATION IN THE HOMOGENEOUS TRAP

The density of atoms in the homogeneous trap is measured using *in situ* absorption imaging [S1]. The absolute atom numbers are calibrated by loading a spin-imbalanced gas into a hybrid trap that is axially harmonic and radially homogeneous [S2]. In Fig. S1, we plot the 1D density profile given by the integrated profile along the two homogeneous directions and the isothermal compressibility $\kappa/\kappa_0 = -\partial E_{F\uparrow}/\partial U$ of the majority component, where $\kappa_0 = \frac{3}{2n_\uparrow E_F}$ is the compressibility of an ideal Fermi gas at density $n_\uparrow$. The compressibility in the spin-polarized region provides the calibration of our measurement of density.

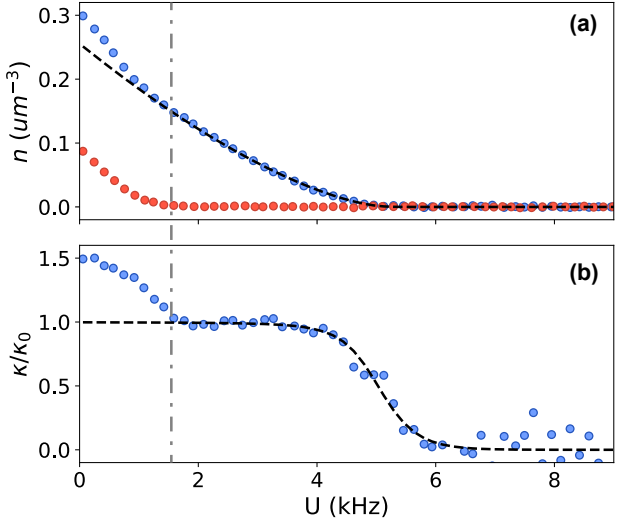

FIG. S1. Density calibration using the spin-imbalanced Fermi gas in the axially harmonic, radially homogeneous trap. Here, the majority Fermi energy is $E_{F\uparrow}/h = 5.7(1)\,\mathrm{kHz}$, $T/T_{F\uparrow} = 0.05(1)$ and the imbalance is $N_\downarrow/N_\uparrow = 0.18$. (a) Majority (minority) density profiles in blue (red) data points. The dashed line is a fit to the ideal equation of state for the spin-polarized Fermi gas, restricted to the polarized wings of the cloud (outside the minority component edge, marked with the dot-dashed line). (b) The isothermal compressibility of the majority component as a function of position.

## RF SPECTROSCOPY MEASUREMENTS

For rf spectroscopy measurements, two images are taken within several $\mu s$ of each other. The first image records the transferred cloud in state $|f\rangle = |2\rangle$, while the second image allows for counting the number of atoms in the initial state $|\downarrow\rangle = |1\rangle$. For measurements of the full spectrum, the pulse time is set to $T_{\mathrm{Pulse}} = 1$ ms, giving a Fourier-limited spectral resolution of 1 kHz. For measurements of the contact, we select a detuning that is large compared to both the Fermi energy and the temperature of the cloud. This ensures that atoms are transferred from the high-momentum tails, and the transfer rate accurately measures the contact. For detunings between $\hbar\omega \approx 5E_F$ and $\hbar\omega \approx 13E_F$, we verified that the measured value of $\tilde{C}$ is constant within statistical errors. The Rabi frequencies are adjusted between $\Omega_R = 2\pi \times 500$ Hz and $\Omega_R = 2\pi \times 1$ kHz to maintain a high signal to noise ratio.

## COOPER PROBLEM AT FINITE TEMPERATURE

In the Cooper problem [S3] one searches for the binding energy of two opposite-spin fermions on top of the filled Fermi sea. The Fermi sea is treated as "inert", its only role being to block momentum states that would otherwise be available to the scattering pair. This constraint alone already leads to pairing in three dimensions. Cooper's solution can be extended to non-zero temperature, in the search of a bound state on top of a finite temperature Fermi gas. A standard approach [S4] yields an equation for the bound state energy $E_c$ for Cooper pairs:

$$-\frac{m}{4\pi\hbar^2 a} = \int \frac{\mathrm{d}^3 p}{(2\pi)^3} \left( \frac{(1-n_F(\xi_p))^2}{2\xi_p - E_c} - \frac{1}{2\epsilon_p} \right), \quad (S1)$$

where $n_F(\epsilon) = (1 + \exp(\epsilon/T))^{-1}$ is the Fermi function, $\xi_p = \frac{p^2}{2m} - \mu$, and $\mu$ the chemical potential of the non-interacting Fermi gas at temperature $T$. The factor $(1 - n_F(\xi_p))^2$ represents Pauli blocking of momentum states already occupied in the spin up and the spin down Fermi sea. Without it, there would be no pairing of two

particles, as is well known in three dimensions. This simplest approach to pairing in a Fermi gas predicts a Cooper pair energy at resonance ($1/a = 0$) of $E_c = -0.61 E_F$ at zero temperature, and an onset of pairing ($E_c < 0$) at $T^*/T_F = 0.41 E_F$. To look for binding in the full many-body framework, one searches for poles of the pair propagator. In the lowest-order T-matrix calculation or equivalently to lowest order in a $1/N$ expansion [S5, S6] (where $2N$ is the number of spin components of the Fermi gas), one finds an equation for this pole that is nearly identical to the above:

$$-\frac{m}{4\pi\hbar^2 a} = \int \frac{\mathrm{d}^3 p}{(2\pi)^3} \left( \frac{(1 - n_F(\xi_p))^2 - n_F(\xi_p)^2}{2\xi_p - E_c} - \frac{1}{2\epsilon_p} \right). \tag{S2}$$

Compared to the simple Cooper problem, the many-body approach yields an additional contribution to the integral from occupied momentum states $\propto -n_F(\xi_p)^2$ as fermions within the Fermi sea now also profit from pairing. This does not change the prediction for the $T = 0$ binding energy $E_c = -0.61 E_F$, but it yields stronger binding at finite temperature, and predicts an onset of pairing at $T^*/T_F = 0.5$. In the main text, we show $E_c$ from this lowest-order many-body approach. As is well-known, fluctuations reduce the onset of superfluidity to lower $T_c$. The next order in the $1/N$ expansion yields [S5] $T_c/T_F = 0.14$, and the self-consistent T-matrix approach [S7] yields $T_c/T_F = 0.16$, in agreement with the experimental value $T_c/T_F = 0.167(13)$ [S8]. However, $T^*$ is often interpreted as the crossover temperature scale for pair formation [S9], and the region between $T_c$ and $T^*$ is the putative "pseudogap" regime. For a recent analysis of pair correlations see [S10].

## CONDENSATE FRACTION

The condensate fraction is measured by performing a momentum space mapping of the pair wavefunction. The atoms are released from the optical box potential into a magnetic harmonic trap with a confining trapping frequency $\omega_z = 2\pi \times 23$ Hz along the z-direction. Simultaneously, the bias field is rapidly ramped from the Feshbach resonance to a value near a zero crossing of the scattering length, which associates existing fermion pairs into bosonic molecules, and preserves the center of mass momentum. Assuming the resulting molecules are non-interacting, the density profile of the cloud after a quarter-period oscillation in the harmonic trap provides the pair center of mass momentum distribution [S2]. The measured integrated profiles $n_{1D}(z)$ are functions of the momentum $k_z = m\omega_z z/\hbar$ along the z direction (see Fig. S2). We fit the wings with the momentum distribution for a thermal gas of non-interacting bosons [S11]:

$$n_{1D}(k_z) = \frac{1}{(2\pi k_B T)^{3/2}} g_{3/2}\left( e^{-|\hbar^2 k_z^2/2m - \mu|/k_B T} \right) \tag{S3}$$

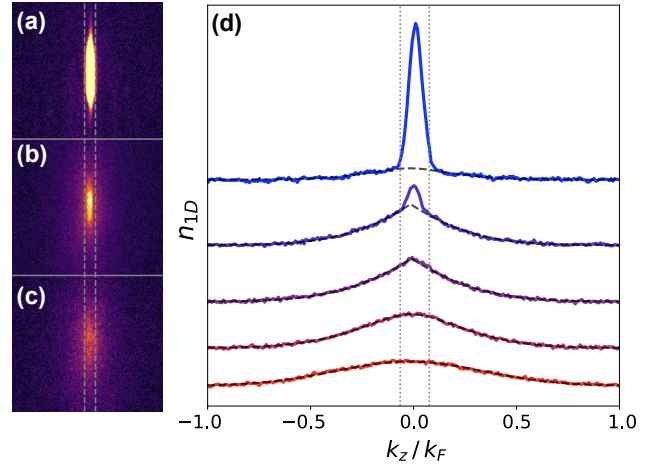

FIG. S2. Momentum space mapping of the box pair wavefunction. (a)-(c) Images of the cloud at $T/T_F = 0.13$ in (a), 0.16 in (b), and 0.18 in (c), after a quarter-period release along the horizontal direction. (d) From top to bottom, $T/T_F = 0.13$, 0.16, 0.18, 0.21, 0.43. Here, $n_{1D}$ is the two-axis integrated pair center of mass momentum distribution, and the dashed black lines are polylogarithm fits to the thermal wings. The dashed vertical lines in (a)-(c) and the dotted vertical lines in (d) mark the condensate region excluded from the fit ($\approx \pm 0.07 k_F$). Here, $k_F$ is the Fermi wavevector in the uniform trap.

As the gas is cooled, the profiles display an increased occupation near zero momentum, and at $T_c$, a clear condensate peak emerges. We define the condensate fraction $N_0/N$ as the difference in area between the observed profile and the fit to the thermal wings.



\end{document}